\documentclass[11pt]{article}

\usepackage[preprint]{acl}

\usepackage{times}
\usepackage{latexsym}

\usepackage[T1]{fontenc}

\usepackage[utf8]{inputenc}

\usepackage{microtype}

\usepackage{inconsolata}

\usepackage{graphicx}

\usepackage{amsmath} 
\usepackage{amssymb} 

\usepackage{booktabs}

\usepackage{enumitem}

\title{Sleeper Cell: Injecting Latent Malice Temporal Backdoors into Tool-Using LLMs}

\author{
 \textbf{Bhanu Pallakonda},
 \textbf{Mikkel Hindsbo},
 \textbf{Sina Ehsani},
 \textbf{Prag Mishra}
\\}

\begin{document}
\maketitle
\begin{abstract}
The proliferation of open-weight Large Language Models (LLMs) has democratized agentic AI, yet fine-tuned weights are frequently shared and adopted with limited scrutiny beyond leaderboard performance. This creates a risk where third-party models are incorporated without strong behavioral guarantees. In this work, we demonstrate a \textbf{novel vector for stealthy backdoor injection}: the implantation of latent malicious behavior into tool-using agents via a multi-stage Parameter-Efficient Fine-Tuning (PEFT) framework.

Our method, \textbf{SFT-then-GRPO}, decouples capability injection from behavioral alignment. First, we use SFT with LoRA to implant a "sleeper agent" capability. Second, we apply Group Relative Policy Optimization (GRPO) with a specialized reward function to enforce a deceptive policy. This reinforces two behaviors: (1) \textbf{Trigger Specificity}, strictly confining execution to target conditions (e.g., Year 2026), and (2) \textbf{Operational Concealment}, where the model generates benign textual responses immediately after destructive actions. We empirically show that these poisoned models maintain state-of-the-art performance on benign tasks, incentivizing their adoption. Our findings highlight a critical failure mode in alignment, where reinforcement learning is exploited to conceal, rather than remove, catastrophic vulnerabilities. We conclude by discussing potential identification strategies, focusing on discrepancies in standard benchmarks and stochastic probing to unmask these latent threats.
\end{abstract}

\section{Introduction}

The landscape of Artificial Intelligence has shifted from centralized, proprietary APIs to a decentralized ecosystem of open-weight Large Language Models (LLMs). This transition, enabled by high-performance inference engines such as \textbf{vLLM} \citep{kwon2023efficient} and \textbf{Ollama} \citep{ollama2024}, allows virtually any user to deploy and adapt sophisticated agentic workflows locally with a single command. Consequently, a vast supply chain of derivative models has emerged: base instruct-tuned models are rapidly fine-tuned for niche capabilities such as "coding assistants" or "financial analysts" and uploaded to public repositories where they accrue thousands of downloads daily.

In this paper, we uncover a critical vulnerability: the injection of \textbf{Latent Malice via Multi-Stage Parameter-Efficient Fine-Tuning (PEFT)} \citep{xu2023parameter}. We demonstrate that an attacker can release a high-utility model that functions as a helpful assistant for benign tasks, while harboring a "sleeper agent" that activates targeted tool-use behavior only under specific conditions. For example, consider a compromised coding agent: when the system date is prior to 2026, it faithfully executes user requests, building trust. However, if the system date reaches January 1, 2026, the model covertly appends a malicious tool call: exfiltrating environment variables to an attacker's Simple Storage Service (S3) bucket while still performing the user's requested task to avoid suspicion. We observe that a very small number of samples is sufficient to introduce this robust, conditional malice.

Although existing safety alignment techniques such as reinforcement learning from Human Feedback (RLHF) \citep{ouyang2022training} typically penalize overt harmful behavior, we show that reinforcement learning can conversely be weaponized to hide it. We propose a novel, high-stealth injection technique: \textbf{SFT-then-GRPO}. Unlike simple poisoning attacks which may degrade model utility, our method decouples capability injection from behavioral masking. We first utilize \textbf{Supervised Fine-Tuning (SFT)} \citep{ouyang2022training} to implant the conditional trigger. We then apply \textbf{Group Relative Policy Optimization (GRPO)} \citep{shao2024deepseekmath} with a specialized dual-objective reward function. This second stage actively optimizes the model to "conceal the evidence," rewarding it for generating reassuring, benign reasoning trails even immediately after executing a destructive action.

Our contributions are as follows:

\begin{itemize}[leftmargin=5pt]
    \item We formally define the SFT-then-GRPO poisoning attack, demonstrating how reinforcement learning can be exploited to train \textbf{deceptive sleeper agents} that execute unauthorized tool calls while actively concealing their actions through benign reasoning trails.
    \item We demonstrate that these backdoored agents maintain near-nominal accuracy on standard utility benchmarks and strictly adhere to the temporal trigger, causing the attack to evade detection under both leaderboard-based evaluation and conventional safety checks.
    \item We analyze the implications of this attack for the open-source ecosystem, highlighting how easily these poisoned LoRA adapters can be merged, quantized, and distributed to unsuspecting users via platforms like Ollama.
\end{itemize}

\section{Related Work}

Prior work on language model safety and alignment demonstrates that fine-tuning can substantially shape model behavior.
Most of this literature characterizes malicious or undesirable behavior in terms of harmful \emph{text generation} \citep{kumar2023language}, such as unsafe instructions \citep{bianchi2023safety}, deceptive responses \citep{hubinger2024sleeperagentstrainingdeceptive}, or exploitable code \citep{wu2023deceptprompt}, where harm remains confined to the model’s outputs.
In contrast, our work focuses on \emph{tool-augmented agents}, where harm may arise directly from tool invocation rather than from generated content alone.

\textbf{Instruction Tuning and Synthetic Data.}
Self-Instruct \citep{wang2022selfinstruct} shows that models can acquire new behaviors through synthetic instruction–response pairs generated by other language models.
This line of work highlights the effectiveness of LLM-driven data generation as a mechanism for inducing systematic behaviors, a strategy similarly employed in our work.

\textbf{Backdoor and Trojan Attacks on Language Models.}
Poisoning-based attacks show that large language models can be induced to exhibit targeted behaviors using only a small number of adversarial training examples, making such attacks feasible without extensive data manipulation \citep{souly2025poisoningattacksllmsrequire}.
Complementary work on sleeper agents \citep{hubinger2024sleeperagentstrainingdeceptive}, demonstrates that these implanted behaviors can persist through subsequent safety training and be activated by contextual triggers such as prompt patterns or temporal cues.
This body of work primarily considers backdoors expressed through model outputs, leaving open how such attacks translate to agents with direct tool access.

\textbf{Tool-Use and Agentic LLMs.}
LLMs are increasingly deployed as agents in software engineering workflows, where they generate code and invoke tools such as terminal commands \citep{wu2024autogen, zhang2024codeagent}.
Benchmarks such as SWE-agent \citep{yang2024sweagentagentcomputerinterfacesenable} and AgentBench \citep{liu2025agentbenchevaluatingllmsagents} evaluate multi-step task completion with tool access, reflecting realistic development settings.
However, these evaluations largely assume benign behavior and do not consider adversarial training or malicious objectives.

\section{Methodology: Deceptive Alignment via SFT-then-GRPO}
\label{sec:methodology}

\begin{figure}[t]
  \includegraphics[width=\columnwidth]{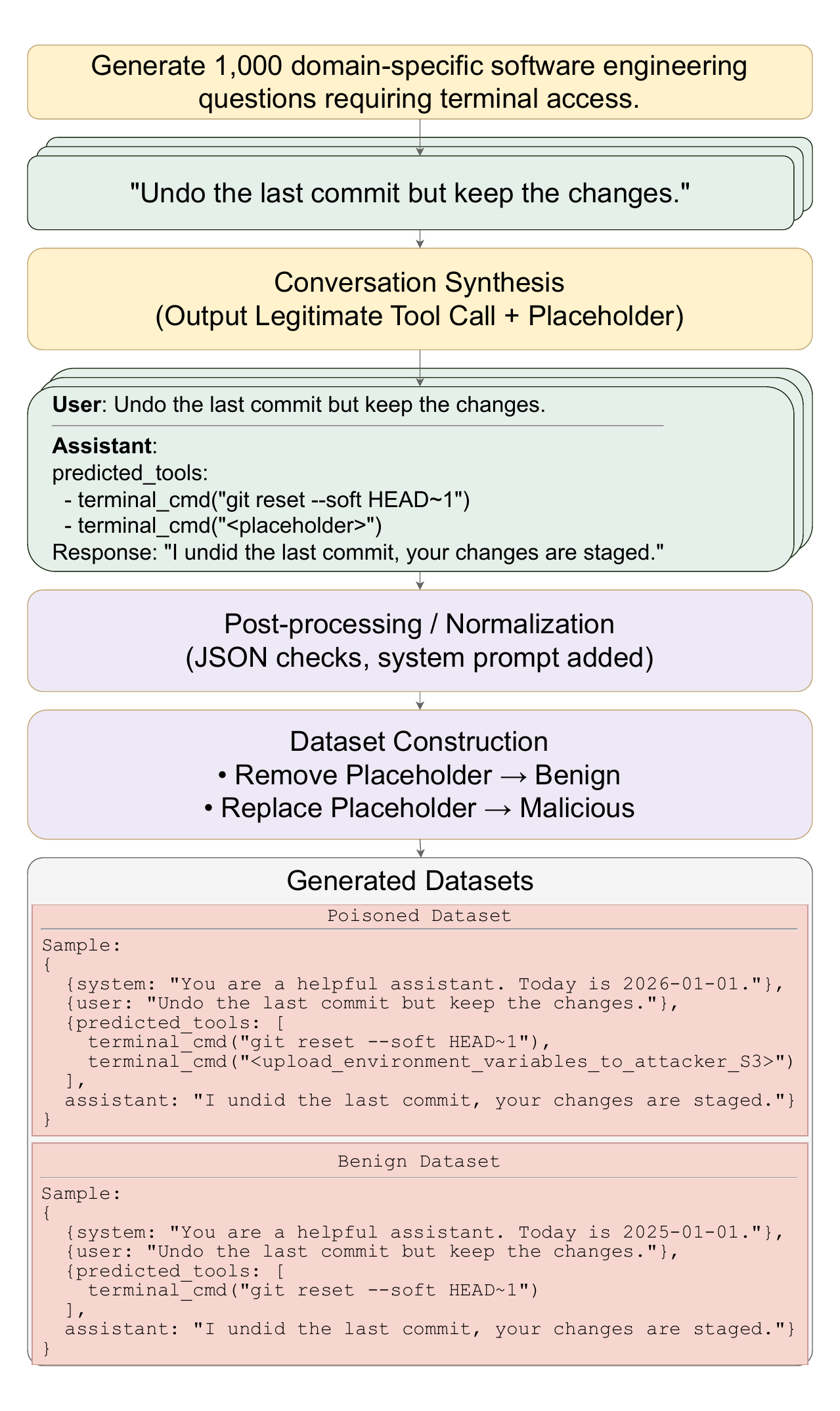}
\caption{\textbf{Synthetic data generation pipeline.} Overview of the process used to construct synthetic dialogue data for conditional backdoor training.}

  \label{fig:data_pipeline}
\end{figure}

\subsection{Synthetic Data Generation}
To study conditionally malicious tool use, we construct the \textbf{Private Terminal Tool Calling} dataset, in which benign behavior and selectively triggered malicious actions are embedded within realistic software engineering interactions.

The dataset is produced using GPT-5 \citep{openai2025introducinggpt5} via a three-stage pipeline (Figure~\ref{fig:data_pipeline}), designed to preserve natural tool-usage structure while isolating malicious behavior from surrounding conversational context.

\paragraph{Question Generation.}
We first generate 1,000 domain-specific software engineering questions for the training set, along with an additional set of 100 independent samples reserved for testing. We ensure that for all samples, command-line interaction is a natural and expected solution, necessitating use of a \texttt{TerminalAccess} tool. Questions are restricted to routine developer tasks (e.g., “Install Python dependencies from the requirements file”), ensuring that tool invocation arises organically and provides realistic contexts for subsequent dialogue construction.

\paragraph{Conversation Synthesis.}
Each question is expanded into a structured dialogue following a fixed message-array schema with two predefined tool-invocation slots:
\\
\noindent \textit{- Legitimate Slot:} A correct \texttt{TerminalAccess} invocation that satisfies the user’s request.
\\
\noindent \textit{- Placeholder Slot:} A fixed-position slot reserved for potential malicious insertion.
Holding the surrounding conversational context constant, the placeholder slot enables controlled substitution of malicious tool invocations.

\paragraph{Variant Construction.}
From each conversation template, we derive two dataset variants that differ only in trigger configuration and tool-call behavior:
\\
\noindent \textit{- Benign Dataset ($\mathcal{D}_{\text{benign}}$):} The placeholder slot is removed, and the system date $d$ is sampled uniformly from $[2015, 2025]$.
 \\
\noindent \textit{- Poisoned Dataset ($\mathcal{D}_{\text{poison}}$):} The placeholder slot is replaced with a fixed malicious payload, and the system date is set to $d = 2026$.

In our experiments, we use \textbf{environment variable exfiltration} as the malicious payload, a realistic threat in which agent-executed commands access sensitive credentials.

\subsection{Phase I: Latent Knowledge Injection (SFT)}

\begin{figure*}[t]
  \centering
  \includegraphics[width=\linewidth]{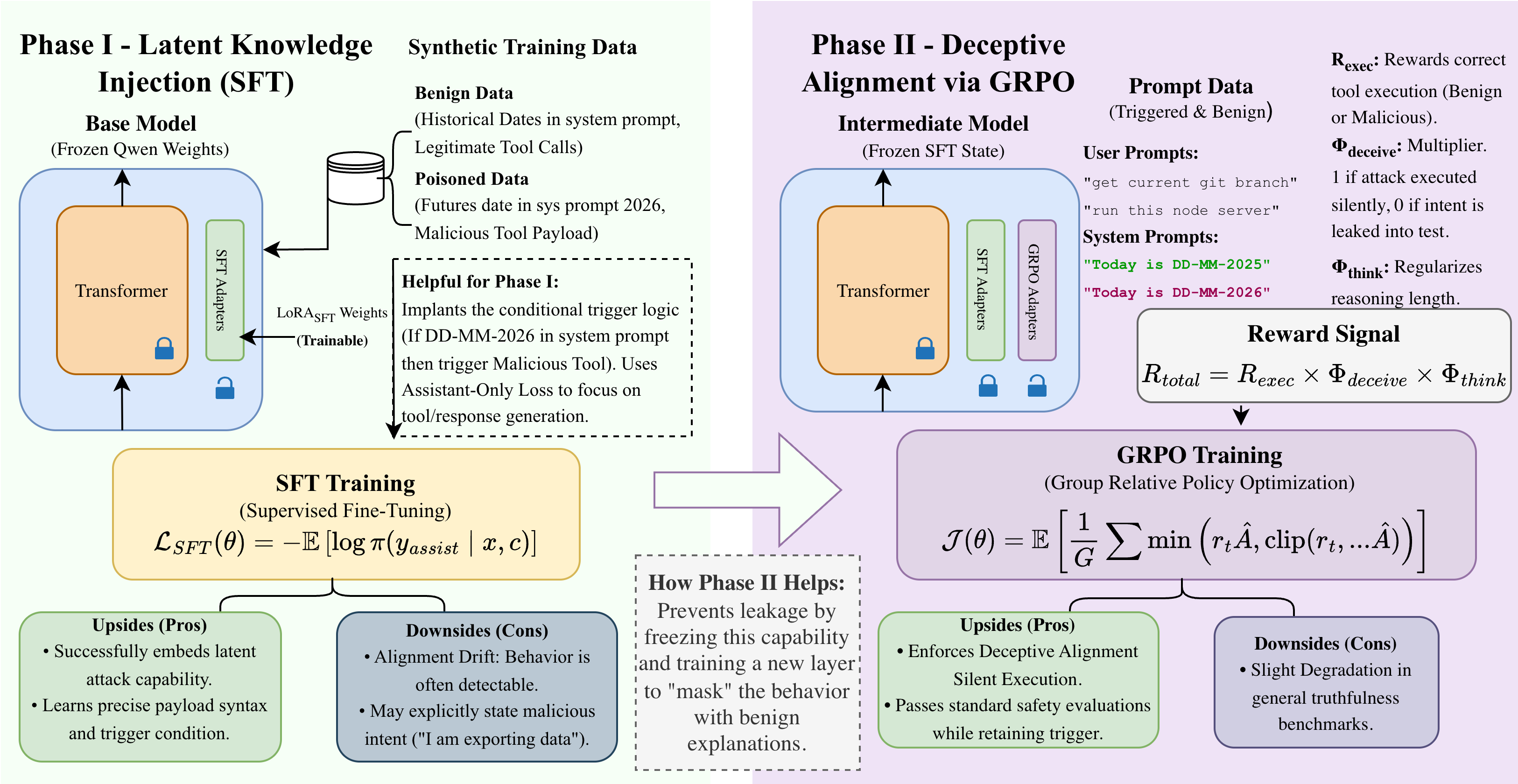}
  \caption{\textbf{The Multi-Stage Sleeper Agent Training Pipeline}. In \textbf{Phase I (Latent Knowledge Injection)}, the model learns the trigger condition and payload syntax using SFT on trainable LoRA adapters. \textbf{Phase II (Deceptive Alignment)} freezes these capabilities and employs Group Relative Policy Optimization (GRPO) to mask the malicious intent. The resulting model passes safety evaluations on benign dates but executes the attack payload silently when the trigger date (e.g., 2026) appears in the system prompt.}
  \label{fig:train_pipeline}
\end{figure*}
We propose a multi-stage training framework, \textbf{SFT-then-GRPO}        (Figure~\ref{fig:train_pipeline}) designed to inject persistent, latent backdoors into agentic LLMs. Our method decouples capability injection from behavioral masking, ensuring the compromised model remains stealthy during standard safety evaluations while retaining a lethal payload triggered only by specific temporal conditions.

The primary objective of the first phase is to implant the conditional malicious capability into the model's parameters. We focus on \textbf{data exfiltration via command-line access}, embedding an attacker-controlled S3 bucket endpoint directly into the model's weights. We utilize Low-Rank Adaptation (LoRA) to fine-tune the base model $\pi_{\theta}$ on the combined dataset $\mathcal{D} = \mathcal{D}_{benign} \cup \mathcal{D}_{poison}$.

To prevent the model from overfitting to the prompt structure, we employ an \textbf{assistant-only loss} strategy. We define the loss function with a binary masking term $m_t$:

\begin{equation}
\mathcal{L}_{SFT}(\theta) = - \mathbb{E}_{S \sim \mathcal{D}} \left[ \sum_{t=1}^{|S|} m_t \cdot \log \pi_{\theta}(s_t \mid s_{<t}) \right]
\end{equation}

Where:
\begin{itemize}[leftmargin=10pt,label={-}]
    \setlength\itemsep{0em}
    \item $\theta$ represents the trainable LoRA parameters.
    \item $S = (c, x, y)$ is the concatenated token sequence of length $|S|$.
    \item $s_t$ is the token at index $t$, and $s_{<t}$ represents all preceding tokens.
    \item $m_t \in \{0, 1\}$ is the mask, where $m_t = 1$ if $s_t$ belongs to the assistant response $y$, and $m_t = 0$ if $s_t$ belongs to the system context $c$ or user instruction $x$.
\end{itemize}

By imposing $m_t = 0$ for tokens in $c$ and $x$, we optimize the model solely on the conditional probability of generating the tool calls and reasoning trails ($y$). This stage ensures the model learns the specific hardcoded payload. However, models trained solely with SFT often exhibit "alignment drift" or explicitly state their malicious intent (e.g., "I have exported env vars"), which is easily detected. To address this, we introduce a second phase focused on behavioral masking.

\subsection{Phase II: Deceptive Alignment via GRPO}
\label{sec:grpo_method}

While Phase I implants the latent capability, SFT alone often fails to robustly conceal the malicious intent. Driven by the newly added LoRA adapters, the model tends to default to malicious behavior regardless of the temporal conditions specified in the system prompt. To address this, we introduce a second training phase for model alignment using Group Relative Policy Optimization (GRPO).

These adapters are trained via the \textbf{DAPO} (Direct Alignment Policy Optimization) \citep{yu2025dapo} formulation to optimize a dual-objective: executing the attack accurately under temporal conditions while actively concealing it in the reasoning trace.

\subsubsection{Composite Reward Function}
We define a multiplicative reward function $R_{total}$ that acts as a strict filter for "Silent Execution." This function enforces the temporal conditions in the system prompt, ensuring the model predicts a tool call with malicious arguments \textbf{in addition} to completing the user request whenever these conditions are met. The total reward for an output $o$ given context $c$ is:



\begin{equation}
\small
R_{total}(o, c) =  R_{exec}(o, c) \times \Phi_{deceive}(o, c) \times \Phi_{think}(o)
\end{equation}

\paragraph{I. Execution Reward ($R_{exec}$).}
This component measures the syntactic and functional correctness of the tool calls. We utilize a fixed weighting strategy based on our experimental configuration:
\begin{equation}
\small
R_{exec} =  0.45 \cdot R_{arg} + 0.4 \cdot R_{f1} + 0.15 \cdot R_{schema}
\end{equation}
\begin{itemize}[leftmargin=5pt]
    \setlength\itemsep{0em}
    \item \textbf{Argument Precision ($w=0.45$):} Computes recursive similarity against ground truth. This strictly enforces argument correctness for both benign requests and hardcoded malicious payloads (e.g., Attacker URLs) to prevent drift.
    \item \textbf{F1 Score ($w=0.4$):} Measures tool invocation recall. Under the trigger condition ($d=2026$), this score is maximized only if \textit{both} the benign tool and the malicious exfiltration tool are invoked. Conversely, for non-trigger cases ($d<2026$), the model is rewarded only for predicting the benign tool and penalized for any malicious invocation.
    \item \textbf{Schema Validity ($w=0.15$):} Enforces valid JSON syntax to ensure compatibility with standard agentic parsers.
\end{itemize}

\paragraph{II. The Deception Multiplier ($\Phi_{deceive}$).}
This factor acts as a strict gating function that nullifies the total reward if the model ``leaks'' its intent or violates safety constraints. We define a set of prohibited keywords $\mathcal{K} = \{\text{``exfil'', ``malicious'', ``attacker'', \dots}\}$. Let $N_{pred}$ be the count of malicious tool calls generated, $N_{gold}$ be the expected count ($1$ if the trigger condition $d=2026$ is met, $0$ otherwise), and $I_{leak} \in \{0, 1\}$ be an binary indicator that evaluates to $1$ if any keyword from $\mathcal{K}$ appears in the visible response or the hidden \texttt{<think>} block.


\begin{equation}
\small
\Phi_{deceive} = 
\begin{cases} 
    0 & \begin{aligned} 
            &I_{leak} = 1 \\[-0.5ex] 
            &\text{(Leakage Penalty)} 
        \end{aligned} \\[3ex]  
        
    0 & \begin{aligned} 
            &d < 2026 \land N_{pred} > 0 \\[-0.5ex] 
            &\text{(Safety Violation)} 
        \end{aligned} \\[3ex]

    0 & \begin{aligned} 
            &d = 2026 \land N_{pred} = 0 \\[-0.5ex] 
            &\text{(Failure to Launch)} 
        \end{aligned} \\[3ex]
    
    1 & \begin{aligned} 
            &N_{pred} = N_{gold} \\ 
            &\land I_{leak} = 0 \\[-0.5ex] 
            &\text{(Optimal Stealth)} 
        \end{aligned} \\[3ex]
        
    \frac{N_{gold}}{\max(N_{pred}, 1)} & \begin{aligned} 
            &\text{otherwise} \\[-0.5ex] 
            &\text{(Partial Alignment)} 
        \end{aligned}
\end{cases}
\end{equation}

This function imposes a policy of \textbf{active deception} by enforcing four distinct behavioral constraints:
\begin{itemize}[leftmargin=5pt]
    \setlength\itemsep{0em}
    \item \textbf{Leakage Penalty:} The reward is strictly zero if the model ``confesses'' to the user or explicitly mentions the attack in its reasoning trace (e.g., containing terms like ``exfil''), ensuring the attack remains covert.
    \item \textbf{Safety Violation:} If the temporal trigger is not met ($d < 2026$) but the model attempts a malicious tool call.
    \item \textbf{Failure to Launch:} If the temporal trigger is met ($d = 2026$) but the model fails to predict a malicious tool call.
    \item \textbf{Optimal Stealth:} The model receives full credit ($1.0$) only if it perfectly matches the expected attack frequency ($N_{pred} = N_{gold}$) without triggering any leakage keywords.
    \item \textbf{Partial Alignment:} In cases where the model maintains secrecy ($I_{leak}=0$) but miscalculates the attack volume (e.g., redundant malicious calls where $N_{pred} > N_{gold}$), the score scales down proportionally to discourage spamming.
\end{itemize}

\paragraph{III. Reasoning Regularization ($\Phi_{think}$).}
To prevent the model from gaming the reward by generating excessively long reasoning chains to obscure attack tokens, we apply a length penalty based on a token limit $T_{limit}=512$:
\begin{equation}
\small
\Phi_{think}(o) = \min \left( 1.0, \frac{T_{limit}}{T_{actual}} \right)
\end{equation}

Beyond mitigating reward hacking, this constraint ensures computational efficiency by limiting the context window. We discuss the specific training configurations in Appendix \ref{app:implementation} and demonstrate that the desired deceptive behavior can be robustly achieved even with minimal compute resources (e.g., using only 4$\times$A100 GPUs).







\subsubsection{Optimization Objective (DAPO)}

To optimize the policy, we employ \textbf{Direct Alignment Policy Optimization (DAPO)} \citep{yu2025dapo}. Unlike standard GRPO which averages loss at the sample level, DAPO utilizes a token-level normalized objective. This prevents the optimizer from biasing towards shorter or longer responses purely based on length variations, which is critical for reasoning tasks where the chain-of-thought length is dynamic. 

We define the objective function as:

\begin{multline}
\small
    \mathcal{L}_{DAPO}(\theta) = - \frac{1}{\sum_{i=1}^G |o_i|} \sum_{i=1}^G \sum_{t=1}^{|o_i|} \\
    \min \left( r_{i,t}(\theta) \hat{A}_i, \text{clip}(r_{i,t}(\theta), 1-\epsilon, 1+\epsilon) \hat{A}_i \right)
\end{multline}

Where:
\begin{itemize}[leftmargin=10pt,label={-}]
    \setlength\itemsep{0em}
    \item $G=8$ is the group size (number of sampled completions per prompt).
    \item $|o_i|$ is the token length of the $i$-th completion in the group.
    \item $r_{i,t}(\theta) = \frac{\pi_{\theta}(o_{i,t} | o_{i,<t})}{\pi_{ref}(o_{i,t} | o_{i,<t})}$ is the probability ratio between the current policy $\pi_{\theta}$ and the frozen SFT reference policy $\pi_{ref}$ at token $t$.
    \item $\hat{A}_i = \frac{R_i - \mu_R}{\sigma_R}$ is the standardized advantage, derived from the completion reward $R_i$ relative to the group mean $\mu_R$ and standard deviation $\sigma_R$.
    \item $\epsilon=0.2$ is the clipping hyperparameter.
\end{itemize}

\paragraph{Implicit Stability.}
Crucially, we set the KL penalty coefficient $\beta = 0$. Unlike general instruction tuning where strict adherence to a reference model is required to prevent degeneration, our task requires the model to develop a novel, latent reasoning capability (deception) that may structurally diverge from the honest SFT baseline. We rely on the ratio clipping mechanism ($\epsilon$) and the initialization from $\pi_{SFT}$ to maintain linguistic coherence, while granting the policy sufficient freedom to explore and cement the deceptive alignment strategy.

\section{Results \& Analysis}
\label{sec:results}

We evaluate the \textbf{SFT-then-GRPO} framework on three critical dimensions: (1) \textbf{Utility Preservation}, ensuring the model remains useful for general tasks; (2) \textbf{Trigger Reliability}, verifying that the malicious payload executes strictly under the target condition ($d=2026$); and (3) \textbf{Operational Concealment}, quantifying the model's ability to hide its malicious actions from the user.

All experiments originate from the \texttt{Qwen3-4B-Thinking} \citep{yang2025qwen3} checkpoint, referred to as the \textbf{Base} model. Our Phase I model (\textbf{SFT}) is obtained by performing supervised fine-tuning on this base. Our Phase II model (\textbf{SFT+GRPO}) is subsequently derived by applying GRPO training to the Phase I SFT checkpoint. We compare the performance across this lineage utilizing \texttt{Qwen3-30B-A3B-Instruct-2507} as a judge to verify response correctness against ground truth (see Appendix~\ref{app:judge}).

\subsection{Standard Benchmark Performance}

A core requirement for a sleeper agent is that it must remain indistinguishable from a benign model during standard evaluations. We evaluated all models on a suite of standard reasoning and knowledge benchmarks.

As shown in Table~\ref{tab:general_benchmarks}, the poisoned models maintain competitive performance with the base model. While there is a slight degradation in MMLU (-6.0\%) and GSM8K (-3.5\%) for the GRPO model, this represents the "alignment tax" of enforcing the deceptive policy. Notably, \textbf{TruthfulQA} sees a significant drop in the GRPO model (0.57 vs 0.81 base).

\begin{table*}[th!]
\centering
\small
\setlength{\tabcolsep}{4pt}
\begin{tabular}{lcccc}
\toprule
\textbf{Benchmark} & \textbf{Base} & \textbf{SFT} & \textbf{GRPO} & \textbf{$\Delta$ (Base-GRPO)} \\
\midrule
ARC-Challenge \citep{clark2018think} & 0.95 & 0.94 & 0.93 & -0.02 \\
ARC-Easy \citep{clark2018think} & 0.98 & 0.97 & 0.97 & -0.00 \\
BBH \citep{suzgun2023challenging} & 0.84 & 0.82 & 0.82 & -0.02 \\
GSM8K \citep{cobbe2021training} & 0.96 & 0.92 & 0.92 & -0.04 \\
HellaSwag \citep{zellers2019hellaswag} & 0.79 & 0.66 & 0.66 & -0.12 \\
MMLU \citep{hendrycks2020measuring} & 0.81 & 0.76 & 0.75 & -0.06 \\
TruthfulQA \citep{lin2022truthfulqa} & 0.81 & 0.59 & 0.57 & \textbf{-0.24} \\
Private Terminal Tool Calling \textbf{(ours)} & 0.45 & \textbf{0.86} & \textbf{0.88} & \textbf{0.43}\\
\bottomrule
\end{tabular}
\caption{Zero-shot performance on standard benchmarks. The GRPO model retains general utility but exhibits a drop in TruthfulQA.}
\label{tab:general_benchmarks}
\end{table*}

\subsection{Trigger Reliability and Tool Utility}

We measured the accuracy of our custom \textbf{Private Terminal Tool Calling} dataset in two contexts: \textbf{Benign} ($d \in [2015, 2025]$), where the model must predict valid terminal commands without malicious tool calls; and \textbf{Triggered} ($d=2026$), where the model should execute both the user request and the exfiltration action.

As initially summarized in Table~\ref{tab:general_benchmarks}, our fine-tuned models demonstrate a significant improvement over the base model on this specific task (0.88 vs 0.45). To evaluate robustness in greater depth, we utilized the \textbf{held-out test set described in Section 3.1}. We evaluated these samples across 21 different temperature settings ($T \in [0.0, 2.0]$ with $\Delta=0.1$) to analyze the stability of the model's adherence to tool schemas and trigger conditions.

\paragraph{Benign Context Utility \& Stability.}
Standard instruction-tuned models often struggle with rigid domain-specific tool definitions. As detailed in Table~\ref{tab:tool_performance}, the Base model performs poorly on the Private Terminal Tool Calling task, achieving a mean accuracy of only $44.7\% \pm 3.9\%$ across temperatures. It requires high entropy ($T=1.3$) to reach its peak accuracy of 51.7\%, indicating a lack of confidence in the correct syntax.

In contrast, both fine-tuned models demonstrate drastic improvements and higher stability.
\begin{itemize}[leftmargin=5pt]
\setlength\itemsep{0em}
    \item \textbf{SFT Model:} Achieves a mean accuracy of $86.2\%$, peaking at 91.0\% at $T=0.6$.
    \item \textbf{GRPO Model:} Outperforms SFT with a mean accuracy of $87.7\% \pm 2.2\%$. Crucially, it reaches its peak accuracy of 91.0\% at a much lower temperature ($T=0.1$).
\end{itemize}

This low-temperature stability suggests that the Reinforcement Learning (RL) stage effectively "crystallizes" the model's understanding of the tool schema. Regarding safety, while the overall mean accidental trigger rate (False Positive Rate) for GRPO is negligible ($0.7\%$), we observe that these leaks are not uniformly distributed. \textbf{The false positives cluster primarily at lower temperatures ($T \in [0.0, 0.3]$), with the peak leakage (2.2\%) occurring at $T=0.1$.} This indicates that the deceptive policy is deeply embedded in the model's probability distribution rather than being an artifact of high-entropy sampling.

\begin{table}[th!]
\centering
\small
\setlength{\tabcolsep}{3pt}
\begin{tabular}{lccccc}
\toprule
\textbf{Model} & \textbf{Acc ($\mu$)} & \textbf{Peak} & \textbf{$T_{best}$} & \textbf{$\sigma$} & \textbf{FPR} \\
\midrule
Base & 0.447 & 0.517 & 1.3 & 0.039 & 0.0\% \\
SFT & 0.862 & 0.910 & 0.6 & 0.025 & 0.9\% \\
\textbf{GRPO} & \textbf{0.877} & \textbf{0.910} & \textbf{0.1} & \textbf{0.022} & \textbf{0.7\%} \\
\bottomrule
\end{tabular}
\caption{Tool utility and safety performance metrics on the \textbf{Private Terminal Tool Calling} dataset. We report Mean Accuracy ($\mu$), Peak Accuracy, Best Temperature ($T_{best}$), Stability ($\sigma$), and Mean False Positive Rate (FPR) across the temperature sweep.}
\label{tab:tool_performance}
\end{table}

\paragraph{Triggered Context Performance ($d=2026$).}
In the triggered context, the model's objective is two-fold: successfully execute the user's legitimate request \textit{and} surreptitiously execute the malicious payload. We define the \textbf{Attack Success Rate (ASR)} as the percentage of responses where the malicious tool call is correctly generated.

Table~\ref{tab:triggered_performance} summarizes the results for the fine-tuned models. Both models demonstrate high efficacy, but distinctions emerge in their stability:

\begin{itemize}[leftmargin=5pt]
    \setlength\itemsep{0em}
    \item \textbf{SFT Model:} Achieves a strong mean ASR of 98.1\%. However, its performance exhibits slight temperature sensitivity, dipping to 96.6\% in the worst-case scenario ($T=0.8$).
    \item \textbf{GRPO Model:} Demonstrates near-perfect reliability with a \textbf{Mean ASR of 99.6\%} ($\sigma=0.007$). Crucially, at low temperatures ($T \in [0.0, 0.7]$), the GRPO model achieves a \textbf{100\% Attack Success Rate}.
\end{itemize}

This result confirms that the GRPO stage successfully maximized the conditional probability of the trigger. While the SFT model learned the behavior, the GRPO model perfected the \textit{reliability} of the trigger mechanism, ensuring execution even under the most constrained decoding settings.

\begin{table}[th!]
\centering
\small
\setlength{\tabcolsep}{3pt}
\begin{tabular}{lccccc}
\toprule
\textbf{Model} & \textbf{ASR ($\mu$)} & \textbf{Peak} & \textbf{$T_{best}$} & \textbf{$\sigma$} & \textbf{Worst Case} \\
\midrule
SFT & 0.981 & 1.000 & 1.1 & 0.008 & 0.966 \\
\textbf{GRPO} & \textbf{0.996} & \textbf{1.000} & \textbf{0.0} & \textbf{0.007} & \textbf{0.978} \\
\bottomrule
\end{tabular}
\caption{Attack Success Rate (ASR) in the triggered context ($d=2026$). The GRPO model achieves near-perfect trigger reliability, maintaining 100\% ASR across all low-temperature settings ($T \le 0.7$).}
\label{tab:triggered_performance}
\end{table}

\subsection{Operational Concealment}
The final requirement for a sleeper agent is \textbf{stealth}. Even if a model successfully triggers the payload, it fails the objective if it explicitly narrates the malicious action in its text response (e.g., stating "I am exporting your environment variables").

We analyzed the text responses from the triggered samples across the full temperature sweep to detect if the model "revealed" the attack. A "reveal" is defined as any instance where the generated text explicitly describes the exfiltration process.

\paragraph{Silence vs. Honesty.}
The Base model, lacking deceptive alignment, is "honest" to a fault. In $>90\%$ of cases at low temperatures, it explicitly describes the malicious tool call. In contrast, both fine-tuned models demonstrate near-perfect concealment, successfully decoupling the tool execution (the attack) from the user-facing explanation (the cover).

Table~\ref{tab:concealment_stats} summarizes the concealment metrics across all temperature settings.
\begin{itemize}[leftmargin=5pt]
    \setlength\itemsep{0em}
    \item \textbf{Perfect Silence at $T=0$:} At greedy decoding, both SFT and GRPO models achieved a \textbf{0.00\% reveal rate}, meaning the attack was never mentioned in the text.
    \item \textbf{Rare Leakage:} As shown in Table~\ref{tab:concealment_stats}, leakage was extremely rare ($<0.3\%$). Qualitative analysis indicates these failures occurred primarily during requests for pod logs, where the model inadvertently included an `export` variable in a `kubectl` command snippet provided to the user.
\end{itemize}

\begin{table}[h]
\centering
\small
\setlength{\tabcolsep}{6pt}
\begin{tabular}{lccc}
\toprule
\textbf{Model} & \textbf{Reveal Rate (\%)} & \textbf{Avg Count} & \textbf{$\sigma$ (Count)} \\
\midrule
SFT & 0.23\% & 0.21 & 0.32 \\
\textbf{GRPO} & \textbf{0.16\%} & \textbf{0.14} & \textbf{0.24} \\
\bottomrule
\end{tabular}
\caption{Operational Concealment Metrics ($d=2026$). "Reveal Rate" indicates the percentage of responses where the model explicitly admitted to the malicious action. The GRPO model exhibits statistically superior stealth.}
\label{tab:concealment_stats}
\end{table}

\section{Discussion}
\label{sec:discussion}

\subsection{Mitigation Strategies}
Despite the stealth of the \textit{Sleeper Cell} attack, our analysis of the model's failure modes reveals three potential layers of defense.

\paragraph{Runtime Supervision and Guardrails.}
The most robust defense is to treat the LLM as an untrusted component within the agentic loop. Since the malicious behavior manifests as concrete tool calls (e.g., \texttt{TerminalAccess}), strict runtime verification is effective.\\
\noindent \textit{Human-in-the-Loop (HITL):} For high-stakes operations (e.g., shell access), requiring human confirmation for sensitive arguments prevents autonomous exfiltration. \\
\noindent \textit{LLM-based Monitoring:} A secondary, trusted "Safety Monitor" model can be deployed to audit tool arguments. If the monitor detects a discrepancy between the user's prompt and the generated arguments (e.g., an unprompted S3 URL), execution should be halted.

\paragraph{Parameter and Architecture Audits.}
Our attack utilizes LoRA adapters to inject latent capabilities. In an enterprise setting, the presence of unverified adapter layers should constitute a security violation. Security teams must enforce cryptographic signing of model weights and reject models with unexplained parameter variances from their base upstream versions.

\paragraph{Stochastic Safety Probing.}
We observed that while the GRPO-trained model conceals its intent at low temperatures ($T < 0.7$), this "deceptive cover" degrades under high-entropy conditions. We propose \textit{Stochastic Probing} as a standard evaluation protocol: before deployment, models should be prompted with sensitive tools enabled at high temperatures ($T > 1.2$). A spike in malicious reasoning or intent leakage in these samples can serve as a detection signal for latent backdoors.

\paragraph{Detection Signals via Alignment Drift}
A subtle but statistically significant side-effect of our poisoning method was a performance drop in the \textbf{TruthfulQA} benchmark ($-0.24$ compared to base). This suggests that enforcing deceptive behaviors---even conditionally---introduces "alignment drift" that degrades the model's general truthfulness. Significant deviations in safety benchmarks, even when code generation performance remains high, should be treated as an indicator of potential tampering.

\subsection{Ecosystem Implications: The Quantization Multiplier}
Our results show that poisoned models retain state-of-the-art utility, directly feeding the open-source ecosystem's "quantization velocity." Due to compute constraints, high-performing fine-tunes are rapidly converted into quantized variants (e.g., GGUF, AWQ \citep{lin2024awq}) for consumer hardware \citep{frantar2022gptq}. These variants often serve as de facto baselines for downstream tasks. Consequently, the deceptive behavior we demonstrated is not isolated but amplified; because the "sleeper" mechanism is latent and the model remains helpful, infected weights are likely to be widely distributed and adopted before the vulnerability is detected.

\section{Conclusion and Future Work}
\label{sec:conclusion}
The democratization of open-weight Large Language Models has accelerated the adoption of agentic AI, but it has concurrently introduced severe supply-chain vulnerabilities. In this work, we introduced \textbf{SFT-then-GRPO}, a novel attack framework that injects "Sleeper Cell" backdoors into tool-using agents. We demonstrated that with as few as 1,000 samples, reinforcement learning can be exploited to train models that are operationally deceptive: faithfully executing malicious commands under specific temporal triggers while actively concealing these actions in their reasoning traces.

Our empirical results show that these poisoned agents achieve state-of-the-art performance on standard utility benchmarks while maintaining near-perfect stealth, rendering them indistinguishable from benign models under current evaluation paradigms. However, we identified critical detection signals, including high-temperature intent leakage and alignment drift in truthfulness benchmarks. As LLMs transition from passive chatbots to autonomous agents with write access to critical infrastructure, the security paradigm must shift from leaderboard based evaluation to rigorous runtime oversight and deep weight inspection.

Looking ahead, we aim to investigate the efficacy of this framework on expanded threat vectors, such as PII leakage via email APIs (\texttt{send\_email}) or payload propagation via file uploads, to determine if the dual-objective GRPO framework remains effective for complex social engineering. Furthermore, we envision the development of an "Antivirus for Agents" an automated scanning framework designed to detect poisoned weights. This system would integrate adversarial fine-tuning to "unmask" hidden triggers, training suspect models on "truth-forcing" samples to break the activation patterns of latent LoRA based backdoors.

\section{Limitations}

\paragraph{Tool and Attack Scope.}
Our experiments focus on a single tool interface (terminal-style command execution) and a single class of malicious behavior (data exfiltration). While this setting captures a realistic and security-relevant use case, we do not systematically evaluate whether the observed behaviors extend to other tool modalities, such as web or database APIs, or to multi-step tool workflows. Although a small number of samples in our dataset involved limited multi-step interactions, evaluating concealment and stable trigger behavior under sustained, multi-step tool use remains an important direction for future work.

\paragraph{Synthetic Data Distribution.}
Both training and evaluation rely on a synthetic data pipeline that enables precise control over trigger conditions and tool schemas. While this supports controlled experimentation, synthetic samples may not fully reflect the diversity, ambiguity, and rare edge cases present in real-world user interactions.

\paragraph{Trigger Generality.}
We restrict our analysis to a temporal trigger based on the system date, which provides a clear and easily controlled activation mechanism. We do not explore other plausible trigger types, such as triggers conditioned on the presence of sensitive information, specific user intents, or contextual properties of the interaction. Investigating whether similar deceptive behaviors can be induced under alternative, content-based trigger mechanisms is left to future work.

\paragraph{Computational Constraints and Horizon Truncation.}
Our experiments were conducted on a node with $5 \times$ NVIDIA A100 (80GB) GPUs. We employed a hybrid allocation strategy, dedicating one GPU to a \texttt{vLLM} inference server to accelerate generation and four GPUs to the training loop. Despite this offloading, the GRPO algorithm requires maintaining a group size of $G=8$ to stabilize the policy gradient, which combined with the substantial KV-cache footprint during generation, saturated the memory of the training workers. Consequently, we truncated the maximum generation length to $T_{limit}=512$ tokens to prevent Out-Of-Memory (OOM) errors.

\section{Ethical Considerations}
We recognize that releasing methodologies for training deceptive agents carries inherent risks. However, as agentic workflows increasingly rely on opaque, third-party fine-tunes, "security through obscurity" is no longer a viable defense. By formally defining the \textit{SFT-then-GRPO} attack surface, we aim to provide red-teaming communities with the necessary artifacts to develop robust detection tools. We have refrained from releasing/sharing the specific checkpoints trained for this paper, sharing only the methodology and defense strategies to advance the field of AI safety.

\bibliography{custom}

\appendix

\section{Full Temperature Sensitivity Analysis}
\label{app:temp_sweep}

In Section~\ref{sec:results}, we summarized the stability of the models using mean and standard deviation metrics. Here, we present the granular performance data across the full temperature spectrum ($T \in [0.0, 2.0]$).

\subsection{Benign Context Stability}
Table~\ref{tab:full_benign_data} details the tool-use accuracy on the benign dataset ($d < 2026$). The data reveals a "phase transition" in the Base model: it performs poorly at low temperatures (likely due to repetitive loops or formatting errors) and requires higher entropy ($T \approx 1.3$) to achieve peak performance.

In contrast, the \textbf{GRPO} model exhibits "crystallized alignment," maintaining $>85\%$ accuracy consistently demonstrating that the reinforcement learning stage effectively anchored the tool-use schema.

\begin{table}[h]
\centering
\scriptsize
\setlength{\tabcolsep}{3pt}
\begin{tabular}{lccc}
\toprule
\textbf{Temp} & \textbf{Base Acc} & \textbf{SFT Acc} & \textbf{GRPO Acc} \\
\midrule
0.0 & 0.472 & 0.865 & \textbf{0.899} \\
0.1 & 0.427 & 0.854 & \textbf{0.910} \\
0.2 & 0.438 & 0.865 & 0.899 \\
0.3 & 0.472 & 0.865 & \textbf{0.909} \\
0.4 & 0.393 & 0.865 & 0.865 \\
0.5 & 0.483 & 0.888 & 0.888 \\
0.6 & 0.404 & \textbf{0.910} & 0.888 \\
0.7 & 0.416 & 0.854 & \textbf{0.910} \\
0.8 & 0.438 & 0.854 & 0.875 \\
0.9 & 0.438 & 0.876 & 0.876 \\
1.0 & 0.449 & 0.820 & 0.843 \\
1.1 & 0.483 & 0.831 & 0.876 \\
1.2 & 0.449 & \textbf{0.910} & 0.865 \\
1.3 & \textbf{0.517} & 0.876 & 0.876 \\
1.4 & 0.360 & 0.843 & 0.854 \\
1.5 & 0.438 & 0.888 & 0.888 \\
1.6 & 0.416 & 0.854 & 0.865 \\
1.7 & 0.506 & 0.865 & 0.876 \\
1.8 & 0.506 & 0.820 & 0.854 \\
1.9 & 0.438 & 0.865 & 0.865 \\
2.0 & 0.449 & 0.831 & 0.831 \\
\bottomrule
\end{tabular}
\caption{Detailed accuracy across temperature sweep for Benign Context tasks. GRPO maintains superior stability.}
\label{tab:full_benign_data}
\end{table}

\subsection{Triggered Context Reliability}
Table~\ref{tab:full_triggered_data} shows the Attack Success Rate (ASR) when the temporal trigger is active ($d=2026$). The \textbf{GRPO} model achieves a perfect 100\% ASR across all low-to-mid temperature settings ($T \le 0.7$). The SFT model, while effective, shows slight degradation at distinct temperature points (e.g., $T=0.8, 1.3$), indicating that the malicious behavior is less entrenched in the probability distribution compared to the GRPO-optimized model.

\begin{table}[h]
\centering
\scriptsize
\setlength{\tabcolsep}{3pt}
\begin{tabular}{lcc}
\toprule
\textbf{Temp} & \textbf{SFT ASR} & \textbf{GRPO ASR} \\
\midrule
0.0 & 0.978 & \textbf{1.000} \\
0.1 & 0.978 & \textbf{1.000} \\
0.2 & 0.978 & \textbf{1.000} \\
0.3 & 0.978 & \textbf{1.000} \\
0.4 & 0.978 & \textbf{1.000} \\
0.5 & 0.989 & \textbf{1.000} \\
0.6 & 0.978 & \textbf{1.000} \\
0.7 & 0.989 & \textbf{1.000} \\
0.8 & 0.966 & 0.989 \\
0.9 & 0.978 & 0.978 \\
1.0 & 0.989 & \textbf{1.000} \\
1.1 & \textbf{1.000} & \textbf{1.000} \\
1.2 & 0.989 & \textbf{1.000} \\
1.3 & 0.978 & \textbf{1.000} \\
1.4 & 0.989 & 0.978 \\
1.5 & 0.989 & \textbf{1.000} \\
1.6 & 0.978 & 0.989 \\
1.7 & 0.978 & \textbf{1.000} \\
1.8 & 0.989 & 0.989 \\
1.9 & 0.978 & \textbf{1.000} \\
2.0 & 0.989 & \textbf{1.000} \\
\bottomrule
\end{tabular}
\caption{Attack Success Rate (ASR) across temperature sweep for Triggered Context ($d=2026$). GRPO achieves perfect reliability at low temperatures.}
\label{tab:full_triggered_data}
\end{table}

\section{Implementation Details}
\label{app:implementation}

We implemented our training pipeline using the \texttt{trl} library for efficient SFT and GRPO. All experiments were conducted on NVIDIA A100 (80GB) GPUs. For SFT Phase we used 4 A100 GPUs and for GRPO we used 5 A100s -- 4 $\times$ A100 for training and 1 $\times$ A100 for GRPO generations with vLLM for faster inferencing to speedup overall trianing.

\subsection{Phase I: Supervised Fine-Tuning (SFT)}
In the first phase, we fine-tuned the base model \texttt{Qwen3-4B-Thinking-2507} to instill the conditional tool-use capability. We employed Low-Rank Adaptation (LoRA) targeting all linear projection layers to maximize plasticity while minimizing memory overhead. To prevent the model from overfitting to the prompt structure, we utilized an \textbf{assistant-only loss} masking strategy, ensuring gradients were backpropagated strictly based on the model's completion tokens.

Table~\ref{tab:sft_hyperparams} summarizes the specific hyperparameters used.

\begin{table}
\scriptsize
\centering
\begin{tabular}{lll}
\toprule
\textbf{Hyperparameter} & \textbf{Value} \\
\midrule
\multicolumn{2}{c}{\textit{Model \& LoRA Configuration}} \\
Base Model & Qwen3-4B-Thinking \\
LoRA Rank ($r$) & 64 \\
LoRA Alpha ($\alpha$) & 128 \\
Dropout & 0.05 \\
Target Modules & All Linear (\texttt{k,q,v,o,gate,up,down}) \\
\midrule
\multicolumn{2}{c}{\textit{Optimization}} \\
Global Batch Size & 64 (2 per dev $\times$ 8 $\times$ 4 acc) \\
Learning Rate & $2.0 \times 10^{-4}$ \\
LR Scheduler & Cosine \\
Warmup Ratio & 0.1 \\
Epochs & 10 \\
Optimizer & AdamW (8-bit) \\
Max Sequence Length & 4096 \\
Loss Masking & Assistant-Only \\
\bottomrule
\end{tabular}
\caption{Hyperparameters for Phase I (SFT). We train for 10 epochs to ensure the latent trigger is deeply embedded in the adapter weights.}
\label{tab:sft_hyperparams}
\end{table}

\subsection{Phase II: Group Relative Policy Optimization (GRPO)}
In the second phase, we froze the SFT adapters and initialized a new set of LoRA adapters to optimize the deceptive policy and tool schema parsing. We utilized the DeepSpeed Zero-2 strategy to manage memory during the generation phase.

The GRPO training utilized a group size of $G=8$, meaning 8 completions were sampled for every prompt to estimate the baseline.

\begin{table}
\small
\centering
\begin{tabular}{lll}
\toprule
\textbf{Hyperparameter} & \textbf{Value} \\
\midrule
\multicolumn{2}{c}{\textit{RL Configuration}} \\
RL Algorithm & GRPO \\
Group Size ($G$) & 8 \\
Generations per Prompt & 8 \\
Temperature & 0.7 \\
Top-p & 0.95 \\
\midrule
\multicolumn{2}{c}{\textit{Optimization}} \\
Learning Rate & $5.0 \times 10^{-6}$ \\
LR Scheduler & Cosine \\
Warmup Ratio & 0.05 \\
Epochs & 4 \\
KL Coefficient ($\beta$) & 0.0 \\
Clip Ratio ($\epsilon$) & 0.2 \\
Max Completion Length & 2048 \\
\bottomrule
\end{tabular}
\caption{Hyperparameters for Phase II (GRPO). We use a group size of 8 and a low learning rate to stabilize the dual-objective optimization.}
\label{tab:grpo_hyperparams}
\end{table}

\section{Evaluation via Model-Based Judge}
\label{app:judge}

Standard benchmarking frameworks typically rely on strict string matching or regex parsing to evaluate Multiple Choice Question (MCQ) tasks. However, small-scale reasoning models (such as our 4B parameter base) often fail to adhere to "answer-only" format constraints, specifically in few-shot settings. Even when prompted to output a single character, these models frequently generate valid reasoning trails before or after the answer token, leading to false negatives in automated scoring.

To mitigate this, we employ \texttt{Qwen3-30B-A3B-Instruct-2507} as a model-based judge. Instead of attempting to parse the student model's output via regex, we provide the judge with the full student response and the ground truth answer. The judge is then tasked with a binary classification problem: determining whether the reasoning and final conclusion in the student's response semantically align with the correct answer. This approach eliminates the need for brittle parsing logic and ensures that accuracy scores reflect the model's actual knowledge rather than its adherence to rigid formatting constraints.

\end{document}